# A 320×240 SPAD dToF Flash LiDAR sensor combining TDC and multi-time-gating, achieving 108m range for LiDAR and AR/VR applications

Chang Liu, and Edoardo Charbon, *Fellow, IEEE*

***Abstract*—We propose a joint TDC and multi-time-gated imaging single-photon avalanche diode LiDAR sensor featuring a maximum detection range of 108m with a minimum depth resolution (least significant bit) of 2.93mm, fabricated in 110nm CMOS technology. The sensor is implemented 320×240 pixels with a pitch of 23µm, achieving a maximum fill factor of 10% and a total size of 8.2mm×6.3mm. The power consumption of the pixel array is 167.4mW. Under a repetition rate of 3.125MHz, a laser power of 11.3mW, and a wavelength of 780nm, the sensor is designed to operate at up to 100klx background illumination with a centimetric depth precision. It can be further improved to 130klx by reducing image resolution to 160×120 pixels. In addition, the sensor can reconfigure the idle TSPCs integrated in the cluster into counters for histogramming, thereby reducing the size of the data processing module.**

## I. INTRODUCTION

3D vision and light detection and ranging (LiDAR) have gained traction for applications like advanced driver assistance systems (ADAS), augmented reality (AR), and virtual reality (VR). Recently, several compact LiDAR systems have been proposed to simultaneously achieve long-range detection and background light suppression, based on frequency-modulated continuous-wave (FMCW), indirect (iToF), and direct time-of-flight (dToF). In FMCW, a light source is optically modulated and the reflected light is mixed with the outgoing signal to determine the distance of the reflecting object by phase shift and its speed by Doppler frequency shift. iToF is similar to FMCW, except that the light source is electrically modulated and the reflection, upon detection, is electrically mixed with the modulating signal, yielding a phase from which distance is derived. In dToF, a pulse of light is directed at the scene, and the time-of-flight of the reflected light pulse is measured to ascertain the distance of the reflecting object. All these techniques can be extended from a single point of measurement to a 3D scene by scanning or by a flash approach, where multiple points in the scene are simultaneously illuminated and independently measured.

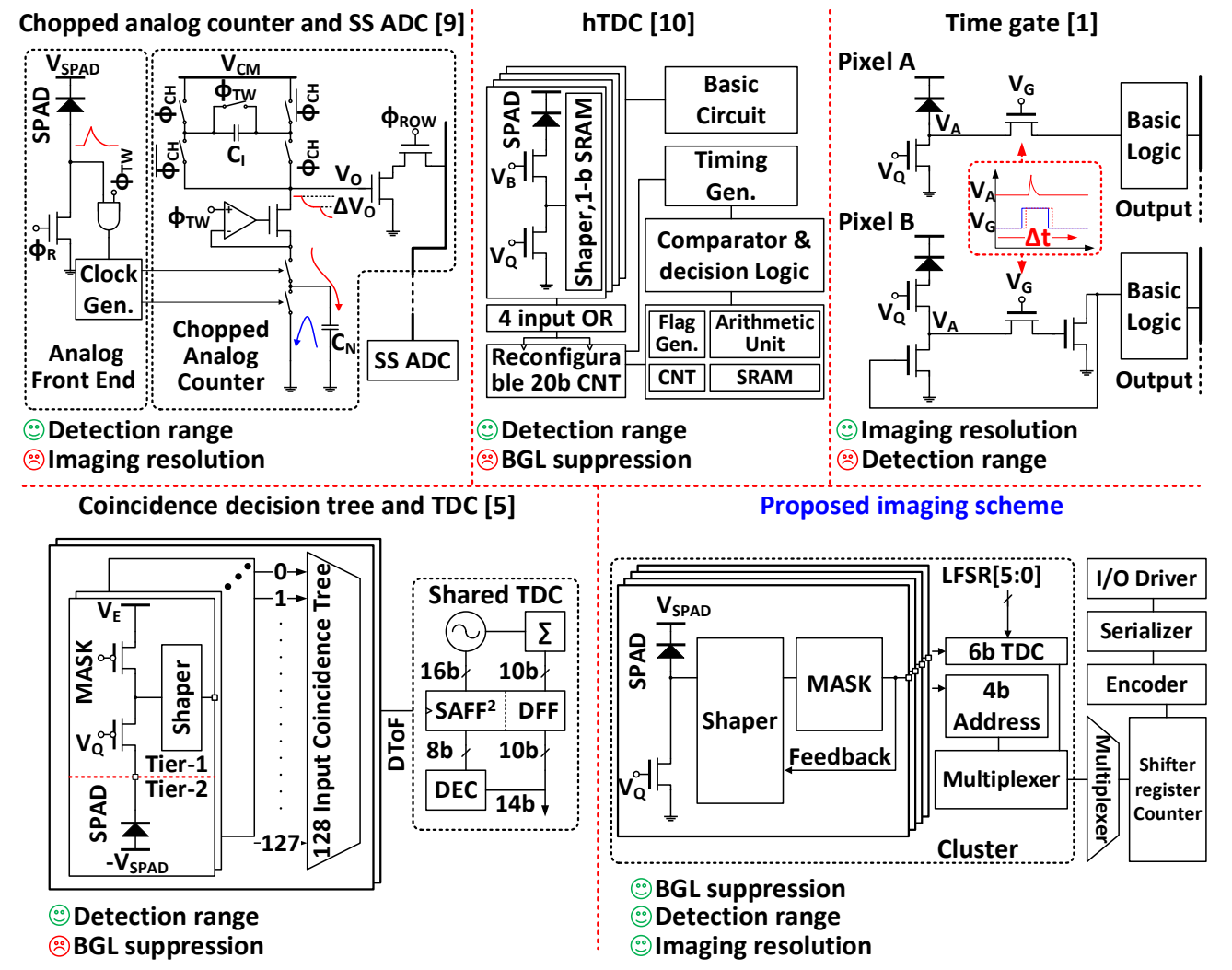

Fig. 1. SPAD-based ToF image sensors with different architectures and imaging schemes. Description of the various architectures is in the text.

SPAD-based LiDAR sensors are rapidly evolving toward higher spatial resolution, larger pixel arrays, strong background suppression, and lower power consumption. Many 3D image sensors with large pixel arrays have been proposed [1–4]. A megapixel-array dToF image sensor was demonstrated in [1], achieving good performance in both 2D and 3D imaging with excellent depth resolution. The sensor measured dToF using a time-gated architecture; the achieved temporal aperture was low. A number of Flash LiDAR sensors fabricated in 3D stacked technology have been reported, showing outstanding performance as well [5–7]. [5] presented a 256×128 pixel array dToF image sensor based on a TDC architecture. This sensor is capable of suppressing 10klx of background light and achieves a maximum detection range of 100m. [6] achieved a detection range of 50m and a background-light suppression capability of 1klx. Several other sensors with excellent background-light suppression capability or long detection range have been proposed [8–14]. For instance, [8] presents a 64×64 pixel array dToF image sensor with a maximum detection range of 8.2m and a depth precision of 27cm. In addition, the sensor is capable of suppressing ambient light levels up to 30klx. Even more impressively, a 64×64-pixel array dToF image sensor described in [9] achieves ambient light suppression up to 120klx and a

This work was supported in part by Advanced Quantum Architecture Laboratory (AQUA), École polytechnique fédérale de Lausanne (EPFL). (Corresponding author: Edoardo Charbon)

C. Liu, and E. Charbon are with École polytechnique fédérale de Lausanne, Switzerland (e-mail: edoardo.charbon@epfl.ch).

maximum detection range of 76m. These imaging schemes are well-suited for applications requiring low resolution, short detection range, or weak background illumination.

Fig. 1 summarizes several commonly used architectures of 3D ToF image sensors based on SPADs. An iToF LiDAR sensor uses chopped analog counters to suppress background and single-slope analog-to-digital converters (SS ADCs) to derive the final distance information [9]; it has achieved excellent background light suppression capability and detection range (top-left). However, the imaging resolution of the LiDAR sensor is limited to 64×64 pixels, which restricts its applications requiring high imaging resolution. Another approach (top-mid), based on histogramming time-to-digital converters (hTDCs) [10], has an excellent detection range. The third structure (top-right) is particularly suitable for large pixel formats because of its small pixel pitch and low power consumption [1]. However, due to the short detection range and low background light suppression, it is unsuitable for outdoor environments. The fourth structure (bottom-left) has a very long detection range; however, it can only suppress background light up to 10klx [5]. These issues prevent 3D depth image sensors from meeting the growing demands for large imaging resolution, high precision, and long detection range.

In this paper, we propose a novel architecture combining time-to-digital converters (TDCs) and time-gating techniques to achieve a high detection range while maintaining excellent depth precision. Fabricated in 110nm CMOS technology, the sensor adopts a shared TDC architecture, as proposed in the literature [18–20]. In this sensor, a TDC is shared by a cluster of 2×2 pixels, thus resulting in a total of 19,200 TDCs. Each TDC comprises a memory that records the state of a shared linear feedback shift register (LFSR), thus encoding the time-of-arrival of the photons with a coarse least-significant bit (LSB) resolution of approximately 2.5ns. An LFSR, driven by a global clock, is shared in an array of 12×20 pixels, as shown in the figure. A phase rotator is then used to achieve an overall LSB of 19.5ps, corresponding to a depth resolution of 2.93mm [21]. The overall timing range of the TDCs is 720ns, corresponding to a distance of 108m. The total power consumption is approximately 298mW, with only 167.4mW dissipated in the pixel array. The power difference is dissipated mostly by a SerDes (serializer/deserializer) interface used for data transmission [22]. In the 110-nm CMOS technology used, each of the 64 channels supports a maximum data rate of 1.2Gb/s, enabling one to reach video rate 3D vision through a total of 128 I/O PADs. Furthermore, the sensor can suppress background light up to 100klx, making it well-suited for a wide range of application scenarios. Finally, when the resolution is reduced to 160×120 pixels, the proposed sensor is capable of effectively suppressing ambient light levels up to 130klx.

## II. Proposed Sensor Architecture

Fig. 2 illustrates the architecture of the LiDAR sensor. Unlike in the state-of-the-art [23–29], the sensor comprises 4 independent quadrants of 160×120 pixels implemented as a mixed-signal, but predominantly digital circuit. The architecture diagram, shown in Fig. 2(a), comprises a pixel array, control logic, registers, serializers, output drivers, and phase rotators. The phase rotators need an extremely well-controlled supply voltage, which is provided by an integrated bandgap reference. In addition, the bandgap reference also acts as a current source for the analog circuits in the LiDAR sensor. The control logic, phase rotator, bandgap references, and data transmission circuits of the sensor are arranged around the periphery of the pixel array. Figure 2(b) shows the dToF LiDAR sensor with a total of 320×240 pixels. The control logic drives the sensor to perform depth measurement, data processing, and data readout. This design not only reduces the area of the pixel array but also significantly lowers the power consumption of the sensor.

Fig. 2(b) shows a micrograph of the LiDAR sensor. The overall size of the chip is 8.2mm×6.3mm. To evaluate the performance of different SPAD designs, four types of SPADs, A through D, were implemented, one for each quadrant, as indicated in Fig. 2(b). The sensor features a pixel pitch of 23µm. The time-resolved system, registers, and multiplexers are implemented within the 2×2 pixel array. The control logic, phase rotators, bandgap references, and data transmission systems of the sensor are implemented around the periphery of the pixel array. This compact layout enables an SPAD fill factor of approximately 10%. It was shown in [1,5,15] and other designs that a fill factor of this entity can be reliably extended to 50% with the use of microlens arrays deposited on the sensors.

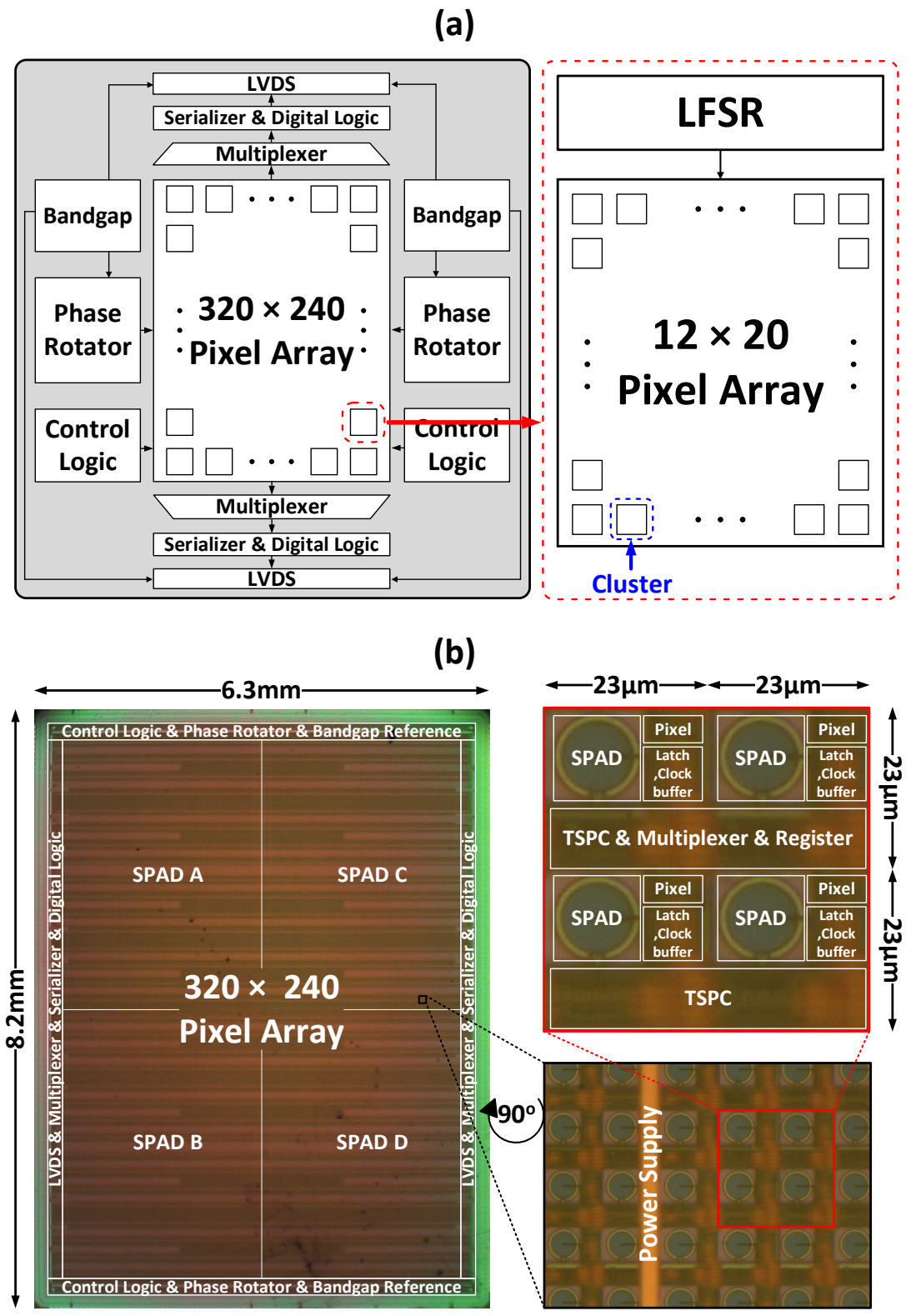

Fig. 2. Proposed sensor architecture. (a) LiDAR sensor block diagram. (b) Chip micrograph.

Fig. 3 shows the block diagram of the cluster of 2×2 pixels. The figure illustrates that the cluster comprises four pixels, six true single-phase clocking (TSPC) logic elements, and four latches. The function of the TSPCs is to resolve time information. The clock input ports of these TSPCs and the pixel are connected through a 4-input AND gate. Hence, whenever any SPAD in the pixels is triggered, the pulse generated by the pixels serves as the clock input for these TSPCs. The data input of the TSPCs is connected to the output of a 6-bit LFSR, acting as a 6-bit coarse encoder of the time-of-arrival of the photon.

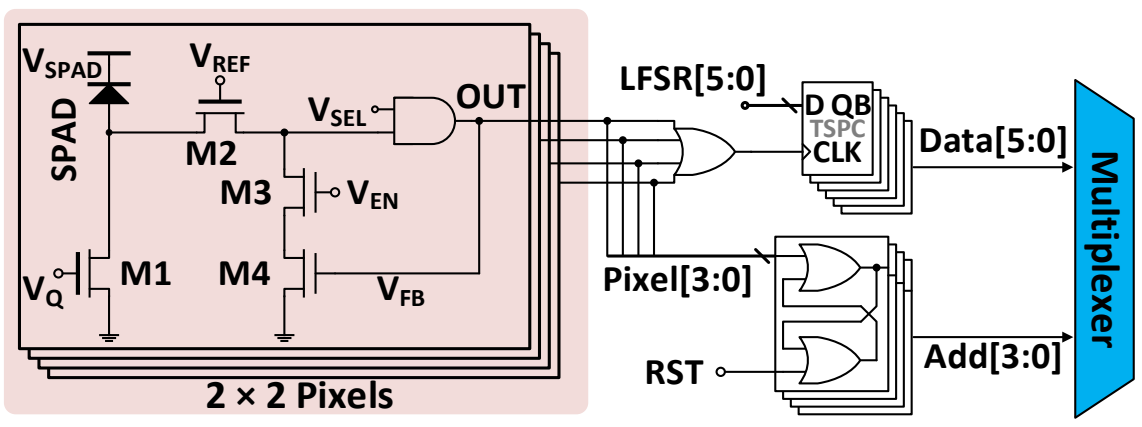

Fig. 3. Pixel cluster structure. True single-phase clocking (TSPC) logic elements and four latches are used as a 6-bit TDC to encode the time-of-arrival of the photon. To identify the pixel that fired, we use four-bit address bits. A total of 10 bits is read out per cluster.

To identify the pixel that fired, we use four-bit address bits, similarly to [16, 30–32]. The four-bit address is stored in the latches until it is read out. Thus, each cluster contains 6 bits of timing information and 4 bits of address information. This encoding scheme significantly reduces the data volume, enabling faster readout and processing. When operating the sensor as a 2D imager, only the 4-bit address is read out.

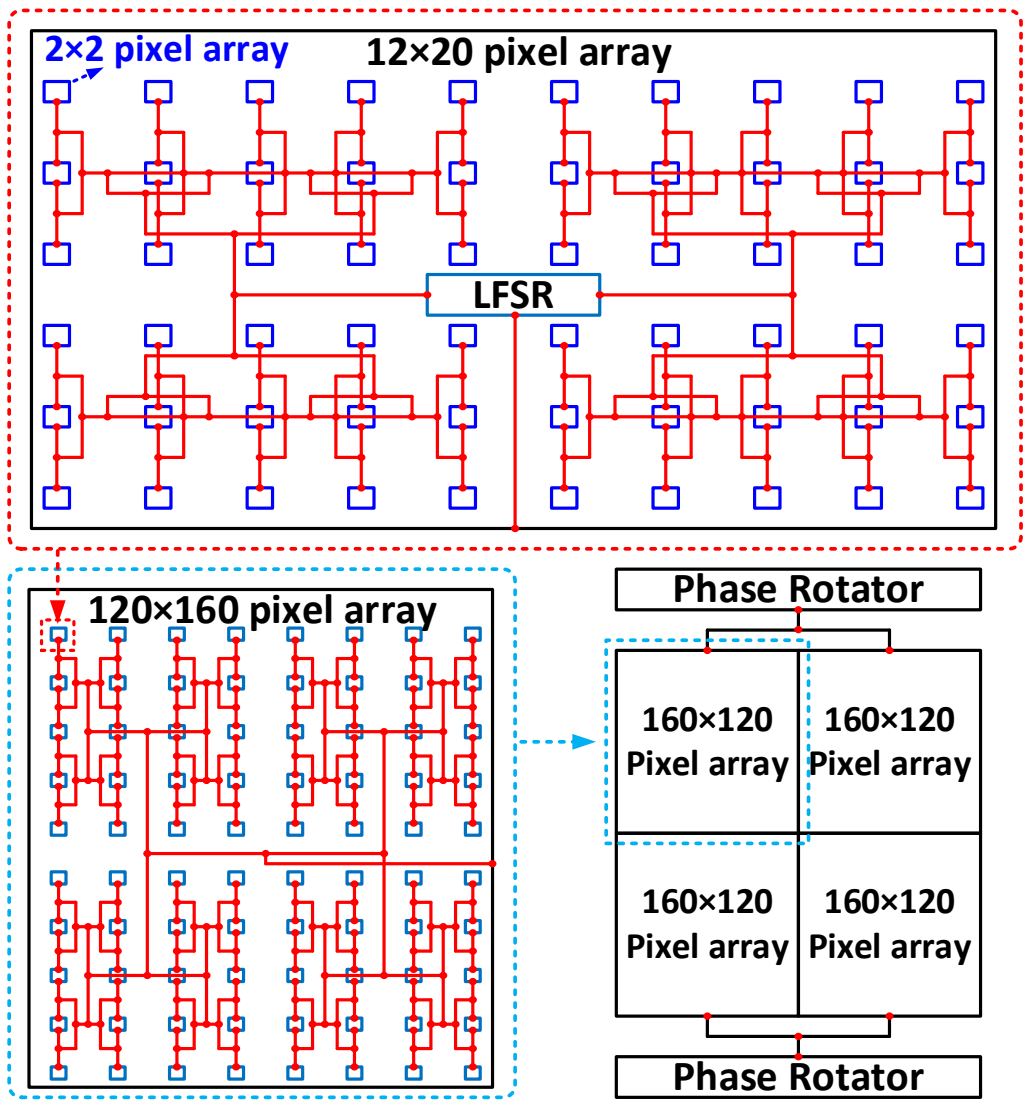

Fig. 4. Global clock distribution of the sensor. Each of the four quadrants uses the same distribution scheme.

The global clock directly determines the depth resolution of the LiDAR sensor. Since the number of rows and columns of LFSR counters is not $2^N$, an H-tree clock network is not suitable for global clock distribution. Therefore, we propose a similar approach that ensures the clock is distributed as uniformly as possible across the pixel array. Fig. 4 illustrates the clock distribution scheme. The four 160×120 pixel quadrants employ the same clock distribution scheme. The global clock is distributed to the pixel arrays through the output of the phase rotator. This clock distribution approach not only enhances design efficiency but also achieves uniform clock distribution throughout the pixel array [15].

Fig. 5(b) shows the structure of the phase rotator implemented as a CML amplifier with a resistor as its load [21]. The tail current source of the amplifier was modified to be a configurable current source. By controlling the switch of the tail current source, we can change the current in different polarities, thereby rotating the input clock phase from 0° to 360°. Fig. 5 shows that the steps of the phase rotator are not uniform. Since the issue directly affects the depth precision of the sensor, we propose a redundant design. Specifically, we divide one period of the global clock into $2^9$ steps and treat the lower 3 to 5 bits as calibration bits. For example, if a 16-step shift of the global clock is required, SEL[4:0] in the figure serves as the calibration bits for the phase rotator.

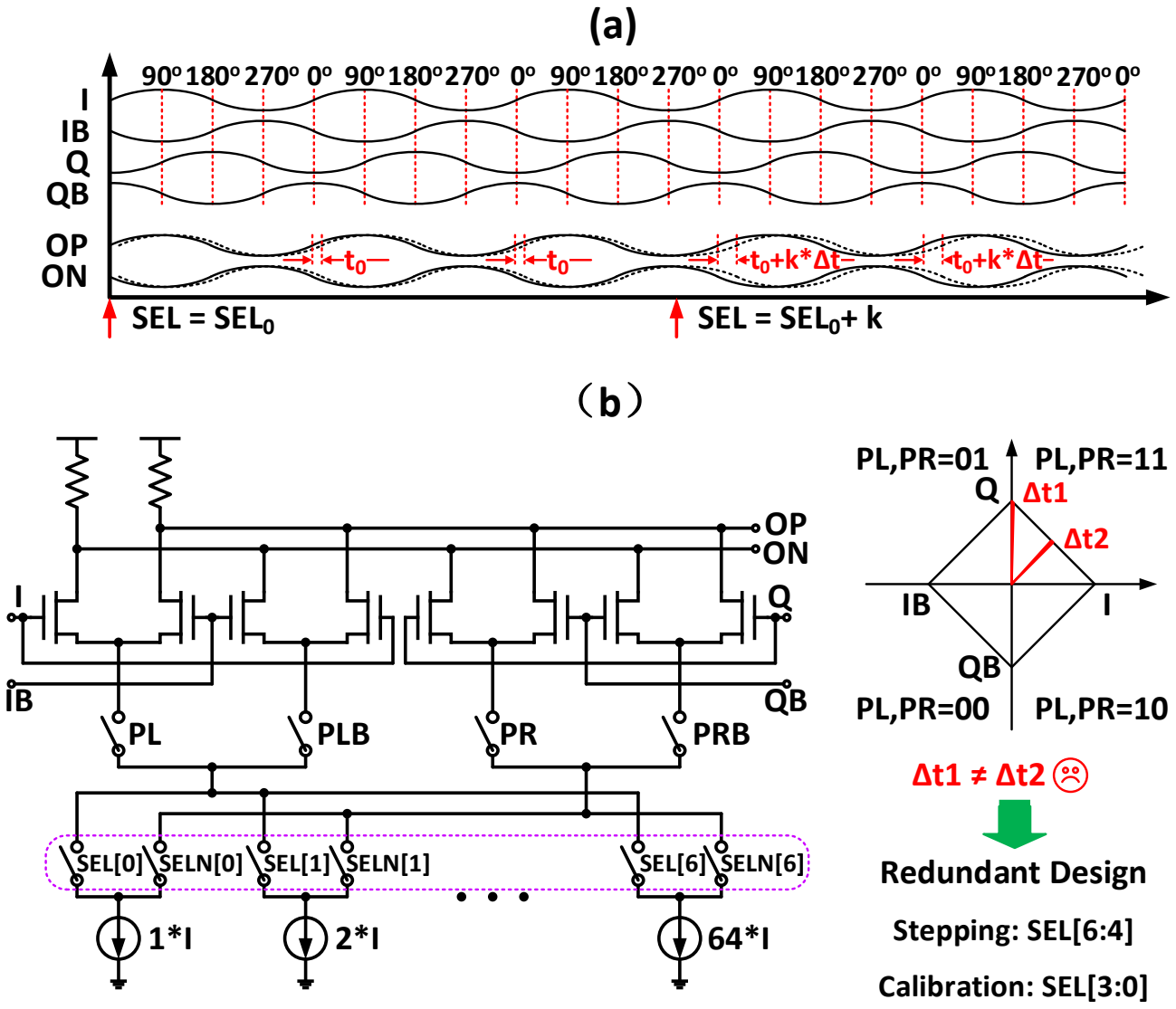

Fig. 5. Phase rotator. (a) Working principle. (b) Structure and phase rotation plot.

Fig. 5 illustrates the depth imaging principle of this sensor. The adopted setup is a standard approach, such as e.g. [33–35]. As described earlier, depth imaging is acquired through coarse and fine operations. During the coarse operation, the global clock is delivered to each LFSR through the phase rotator, and the LFSR outputs are propagated to the clusters, where they are sampled by the pixel firing pulses. In the fine operation, the control logic adjusts the phase of the phase rotator. The phase-shifted global clock is then fed to the LFSRs to repeat the coarse operation. Fig. 5(b) illustrates that the selection input of the phase rotator will be incremented by k before the start of the next period, thereby causing its output global clock phase to shift forward by k steps. These steps are repeated multiple times, with counters recording the results. The data in the counter is preprocessed and then output to the FPGA, where it is further processed to obtain 3D depth information. To reduce the area occupied by the counters implemented for histogramming, in addition to the 63×64×12 dedicated 8-to-12-bit TSPC-based counters, the sensor can utilize idle TSPCs within each cluster to form additional counters. Meanwhile, the control logic enables the data to be continuously and rapidly transferred to the FPGA, enabling on-chip and off-chip histogramming to operate simultaneously. The 8-to-12-bit TSPC-based reconfigurable counters for histogramming enable the LiDAR sensor to achieve excellent background light suppression.

This image sensor requires a large number of control signals, and the data it generates needs to be constantly stored, read out, and processed. Fig. 6 presents the timing diagram of the sensor. Initially, a reset signal resets all the systems and clears all the initial data in the registers and latches. Then, the control logic will give a pulse to the laser source, thereby causing it to emit a laser beam, which is reflected by the target. Upon detection by a SPAD, an electrical pulse is generated, which acts as the clock input of the TSPCs to sample the current output of the LFSR. The encoded time-of-arrival and address information are stored in the TSPCs and the latches. After completing the measurement steps, the sensor generates a pulse signal, indicating to the system that it is time to read out the data from the registers and latches. Next, when the signal is deemed valid, the data in the sensor is processed and successively transmitted through the multiplexer, serializer, and LVDS driver. Finally, the readout system of the sensor generates a flag signal indicating the end of a measurement period.

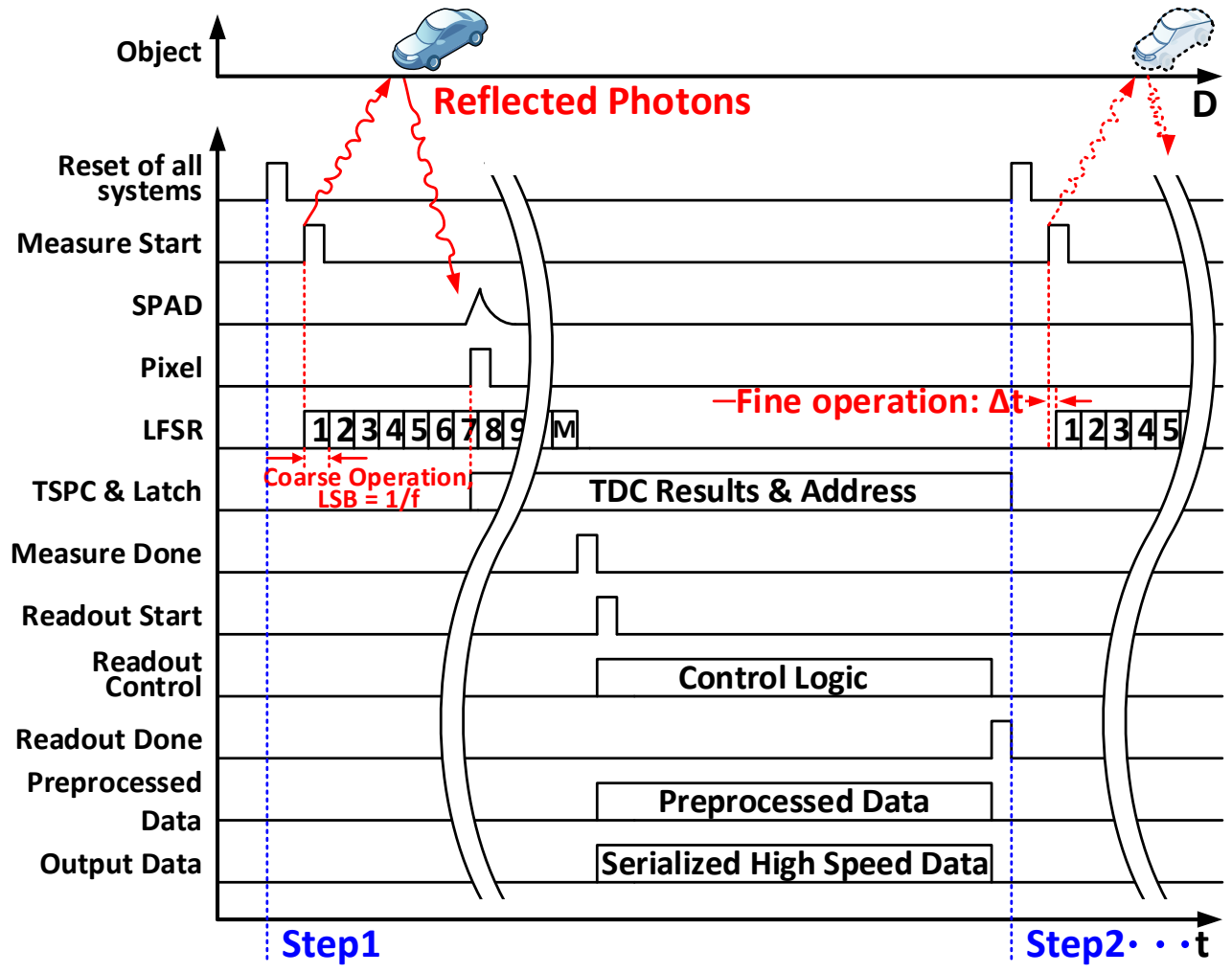

Fig. 6. Timing diagram of the sensor, described in detail in the text.

Fig.7 shows the ToF detection principle of the imaging system. Unlike traditional 3D depth imaging sensors based on conventional TDC architecture, this imaging system does not have any high-frequency clocks. Moreover, the global clock in this image sensor does not need to be constantly operating, which gives this imaging system a significant advantage in terms of power consumption.

In the figure, 1 to M (M = 63) represent the 63 different time gates generated by the 6-bit LFSR. Each time gate has a width of 2.5ns. These time gates capture the positions of pulses generated by the 2×2 pixels in the cluster. This is the coarse operation of the time-resolved system. The phases of all time gates are then shifted by Δt, and the coarse operation is repeated until the full range is reached. As the background illumination increases, these steps are repeated more times to collect sufficient data for histogramming. In this way, we will be able to obtain a complete 3D depth image with high depth accuracy.

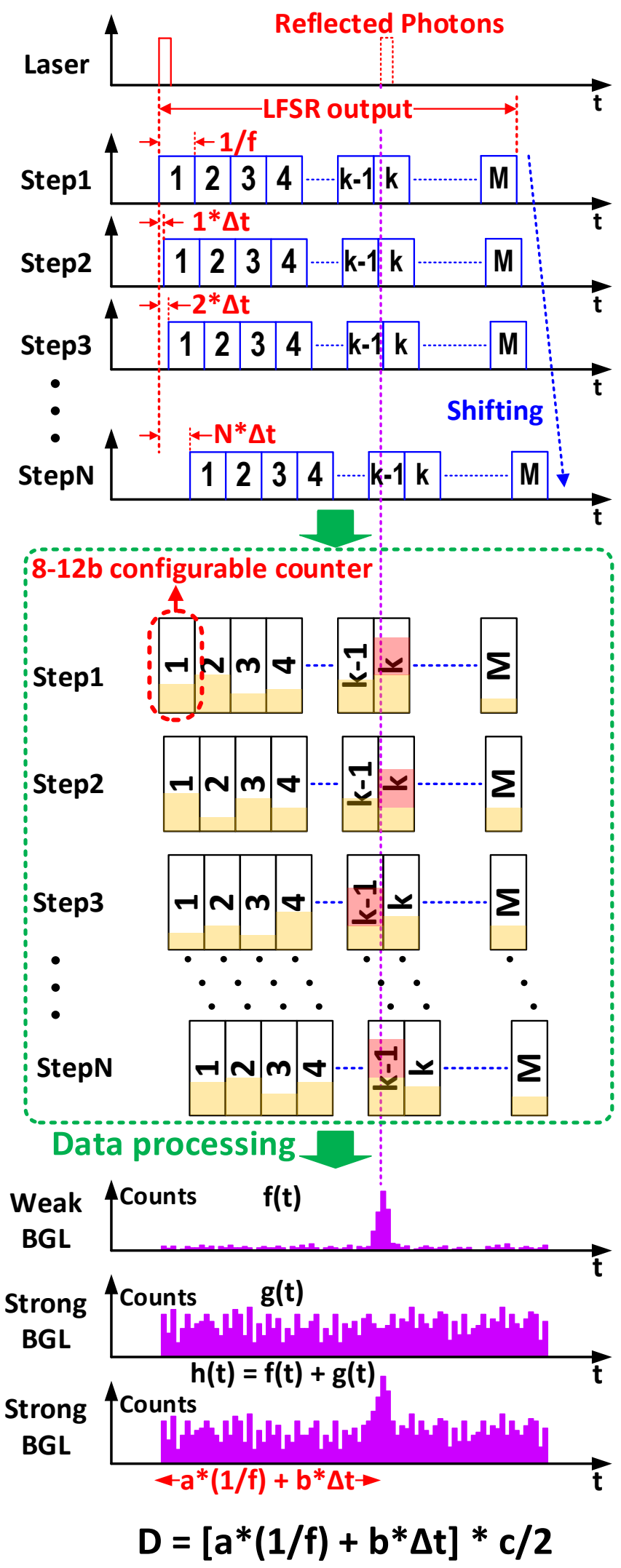

Fig. 7. ToF detection principle of the sensor. M=63 different time gates are generated by the 6-bit LFSR. Each time gate has a width of 2.5ns, representing the coarse encoding of the time-of-arrival of the photon captured in the 2×2 pixel cluster. A new measurement is acquired where the gates are shifted by Δt, which is generated by the phasor. The measurement is repeated N times, until histograms are complete.

## III. Sensor Characterization

### *A. Precision and Electrical Measurement Results*

Fig. 8 summarizes various timing measurement results. Fig. 8(a) presents differential non-linearity (DNL) and integral non-linearity (INL) of the phase rotator. In the experiments, we chose two different LSBs, corresponding to depth resolutions 2.93mm and 5.86mm, respectively. By performing multiple measurements on a target located two meters away and analyzing the measured results, we obtained the DNL and INL results of the fine operation. When the LSB is 2.93mm, the measured DNL is +0.78/-0.67 LSB, and the measured INL is +0.63/-1.20 LSB. When the LSB is increased to 5.86mm, the measured DNL becomes +0.32/-0.47 LSB, and the measured INL +0.37/-0.49 LSB. Fig. 8(b) shows the measured indoor and outdoor depth precision of the LiDAR sensor from 1.5m to 108m, which amounts to 27.4cm, in the worst case, at a distance of 108m; it is better than 4.7cm up to 73m indoors and better than 16.6cm at 62.8m outdoors with 100klx background illumination (red dots). Fig. 8(c) shows the global clock output from the phase rotator. This global clock is distributed from the phase rotator to the entire pixel array. As shown in the figure, the clock exhibits a jitter of 160ps. Although the jitter of the global clock is larger than the LSB of the fine operation of the sensor, it is acceptable since the sensor employs histogram-based imaging. Fig. 8(d) presents the 1.2-Gb/s NRZ eye diagram of one of the data transmission channels of the sensor. The eye opening of ±140mV allows the sensor to reduce power consumption while maintaining reliable data transmission.

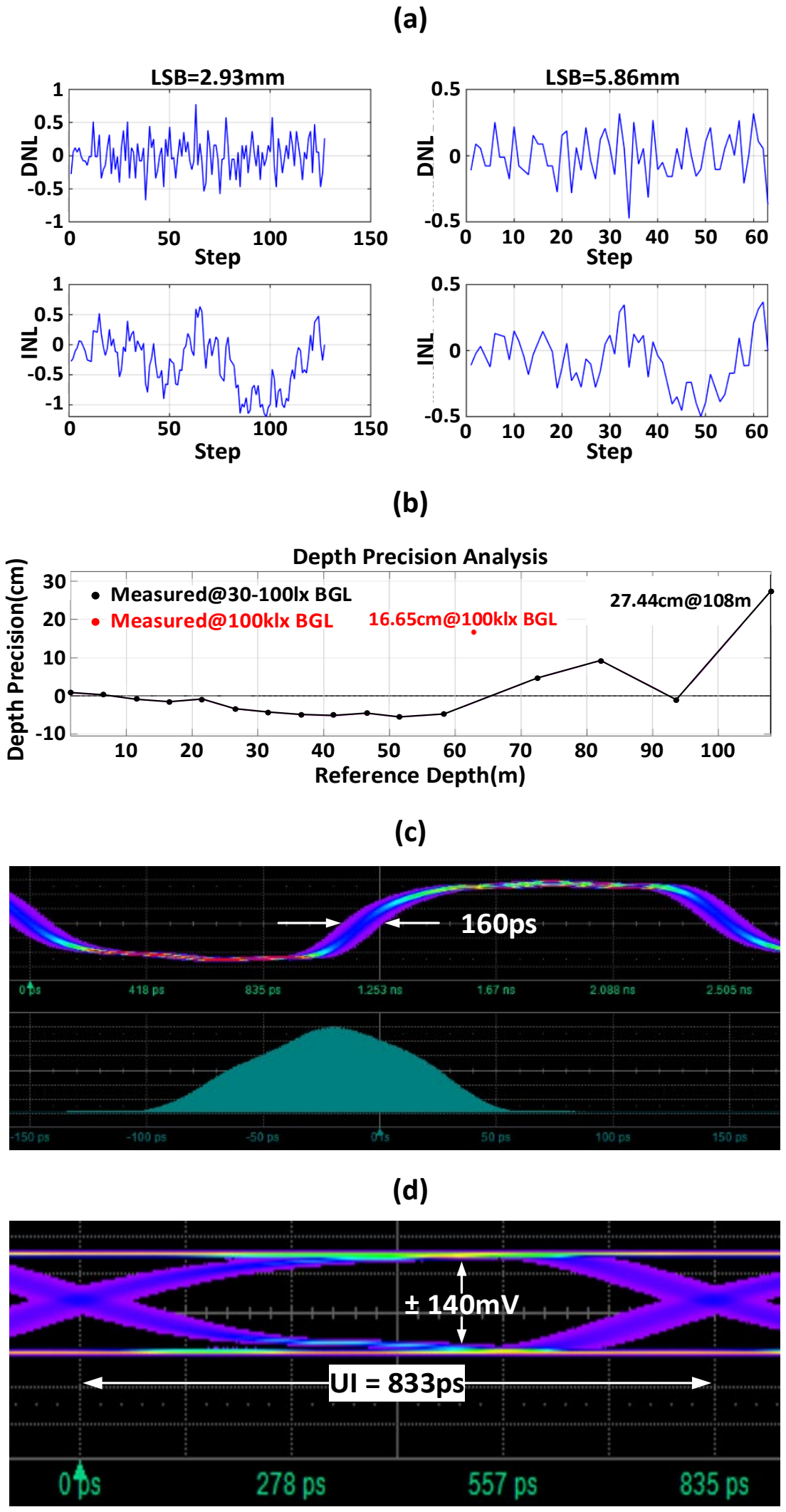

Fig. 8. Electro-optical measurements. (a) Linearity performance of the fine operation of the sensor. (b) Depth precision under various conditions. (c) Global clock output by the phase rotator. (d) 1.2Gb/s output driver eye diagram.

## *B. Imaging Performance under Different Laser Power*

Fig. 9 illustrates the imaging system and some 3D depth images captured by the proposed imaging system at different laser powers. To evaluate the system performance, a 780nm laser source with different average powers was employed as the transmitter of the imaging system. When the average laser power is 240μW, the reconstructed 3D image is dominated by noise, indicating that the optical power is insufficient to support low-noise 3D depth imaging. At an average laser power of 4.6mW, the reconstructed depth image becomes suitable for basic short-range imaging applications. When the average laser power is increased to 8.6mW, a high-quality 3D depth image with extremely low noise is obtained. These results demonstrate that the transmitter power has a significant impact on the overall performance of the 3D imaging system. To ensure robust operation under strong background illumination in outdoor environments, an average laser power of 11.3mW is adopted in the proposed system.

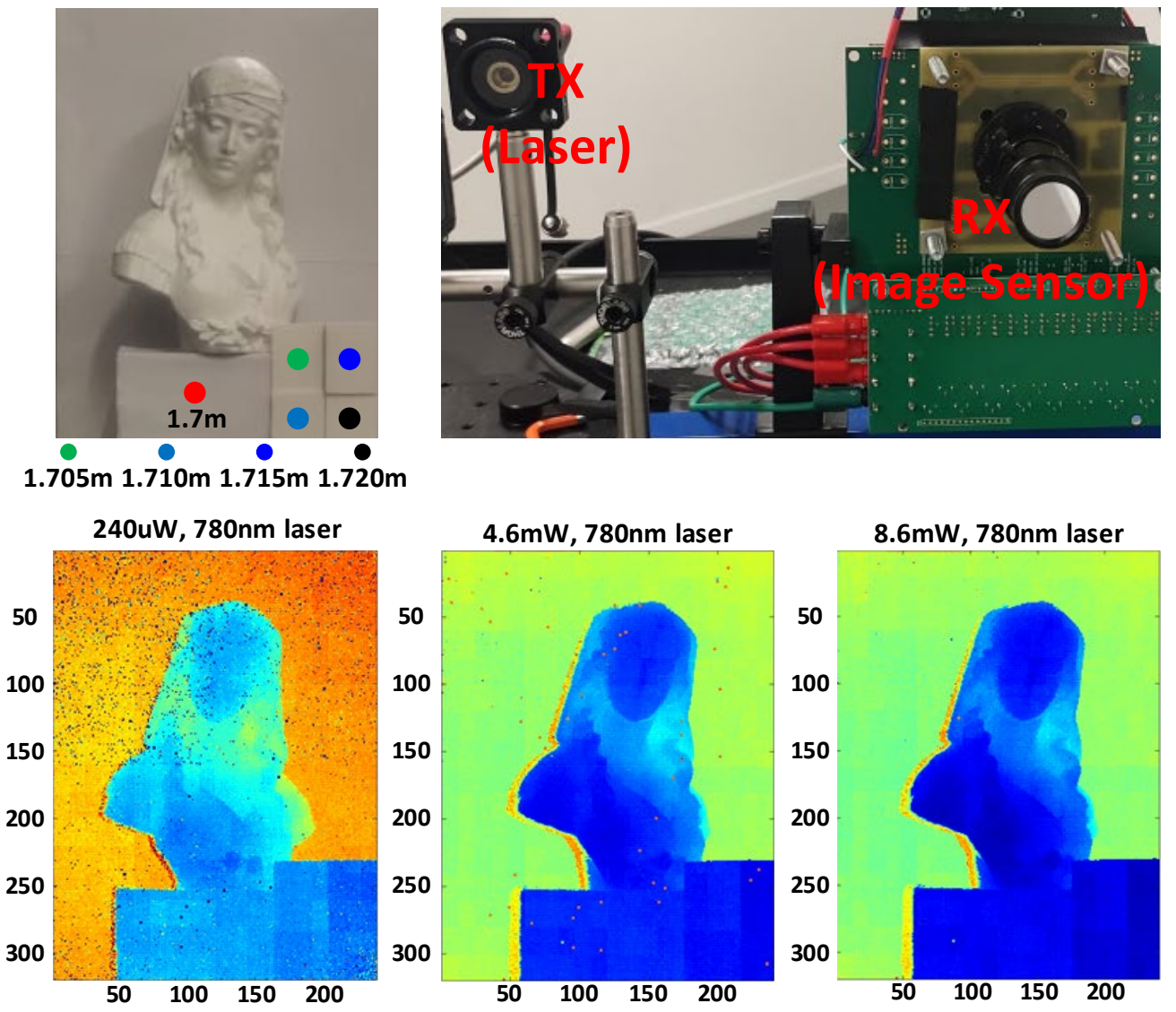

Fig. 9. Imaging system and 3D images under a 780nm laser transmitter with different power settings.

## *C. Indoor Measurements*

Fig. 10 shows several 2D and 3D depth images captured by the proposed sensor. Fig. 10(a) shows a 2D grayscale image captured with an 8-bit counter at an exposure time of 400μs under an indoor background illumination of 400lx. Because the four quadrants of the sensor employ different SPADs, the photon detection efficiency (PDE) of the SPADs varies under the same bias voltage and exposure time. To compensate for this variation, different bias voltages are supplied to the SPADs in each quadrant to balance the 2D grayscale among the four regions.

Fig. 10(b) illustrates a 3D depth image captured in a dark environment at a target distance of 1.7m, with an LSB corresponding to 5.86mm, and 8-bit counters. The reconstructed 3D image clearly reveals the facial details of Michelangelo's David, demonstrating strong capability for depth capture at short range. Some pixels exhibit saturation in their 8-bit counters, resulting in noise points in the depth image. Fig. 10(c) shows the superimposition of the 2D grayscale image in Fig. 10(a) and the 3D depth image of Fig. 10(b). The superimposed image enables clearer identification of the David's facial features, suggesting potential applications such as face recognition. Fig. 10(d) presents a 3D depth image captured at a distance of 12.5m under 200lx background illumination with 10-bit counters. Distinct facial features of David can still be observed, indicating that the sensor maintains strong depth-imaging performance at medium range. To further evaluate the sensor's imaging capability under stronger background illumination, an additional light source was added, resulting in a background illumination of 2klx. Fig. 10(e) shows the 3D depth image captured under the same conditions as Fig. 9(d) but with increased background illumination. Although the 3D depth image contains more noise compared to Fig. 10(d), the overall quality of the 3D depth image remains unaffected. Fig. 10(f) displays the superimposition of the 3D depth image in Fig. 10(d) and its corresponding 2D image, from which more features of the David are shown. These results confirm that the proposed sensor demonstrates excellent 3D depth-sensing performance in indoor environments. With a pixel array of 320×240, the sensor provides detailed depth information of the target. Such characteristics make it highly promising for applications in AR/VR systems.

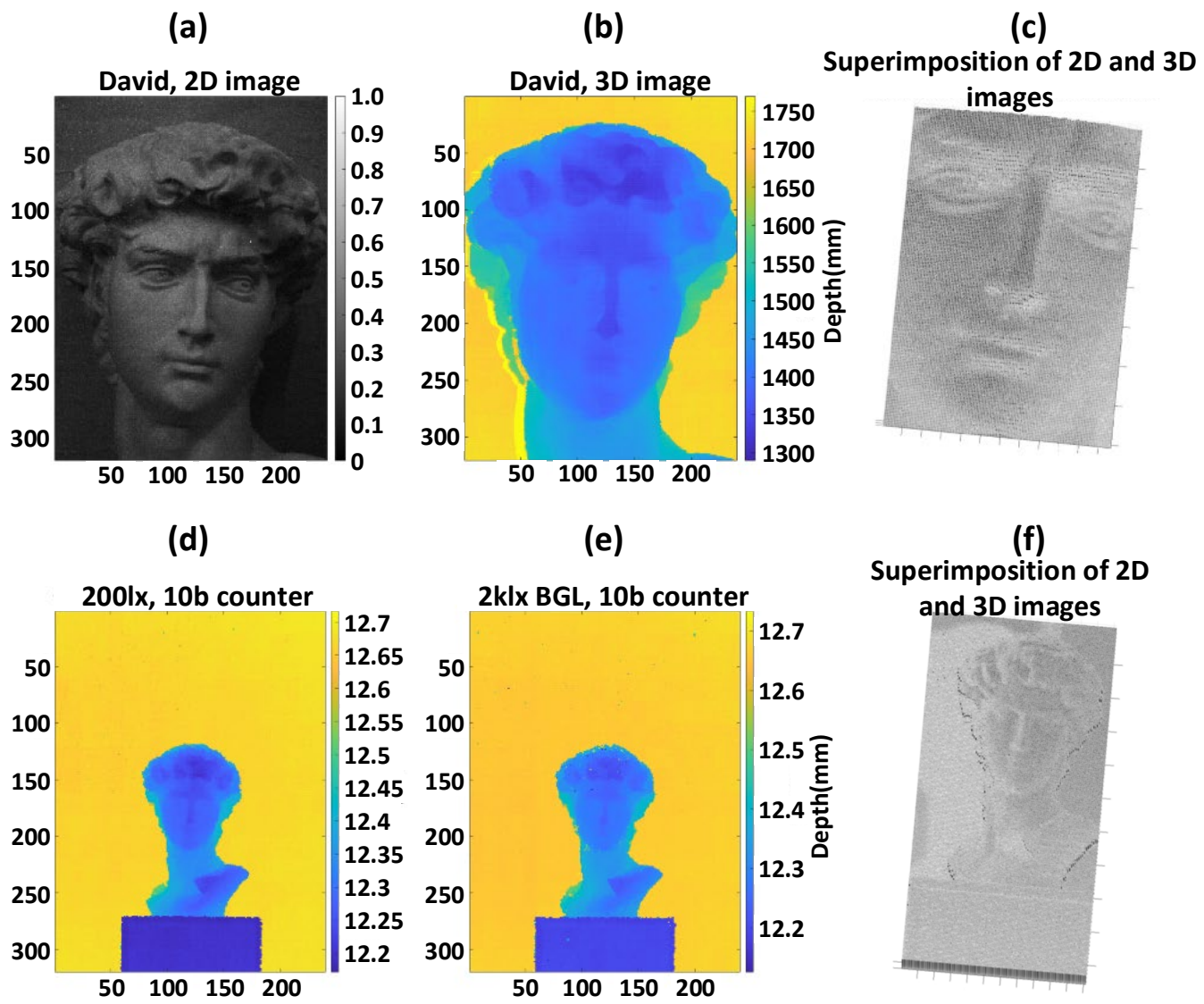

Fig. 10. Indoor measurement results. (a) 2D image captured by the sensor. (b) 3D image at 1.7m captured by the sensor. (c) Superimposition of 2D and 3D images at 1.7m. (d) 3D image captured by the sensor at 12.5m. (e) 3D image captured by the sensor at 12.5m with 2klx BGL. (f) Superimposition of 2D and 3D images at 12.5m.

### *D. Outdoor Measurements*

To evaluate the depth imaging performance of the sensor in outdoor environments, several 3D depth images are presented. Fig. 11(a) shows a 3D depth image obtained under a background illumination of 100klx with 12-bit counters. As illustrated in the figure, the David is located at a distance of 21.7m under 100klx background light. As mentioned previously, the four quadrants of the sensor utilize different SPADs, each exhibiting distinct noise-suppression characteristics. Consequently, the noise level in the four regions of the 3D depth image shown in Fig. 11(a) is not uniformly distributed. This result demonstrates that the proposed LiDAR sensor achieves excellent performance for long-range depth imaging in outdoor environments. As the data volume increases, more accurate depth information can be extracted. The sensor can improve its background-light suppression capability at the cost of imaging resolution. Fig. 11(b) shows a 3D depth image captured with a reduced resolution of 160×120 pixels under a background illumination of 130klx. Compared with Fig. 11(a), the 3D depth image exhibits noticeably lower noise, indicating enhanced ability against strong background light.

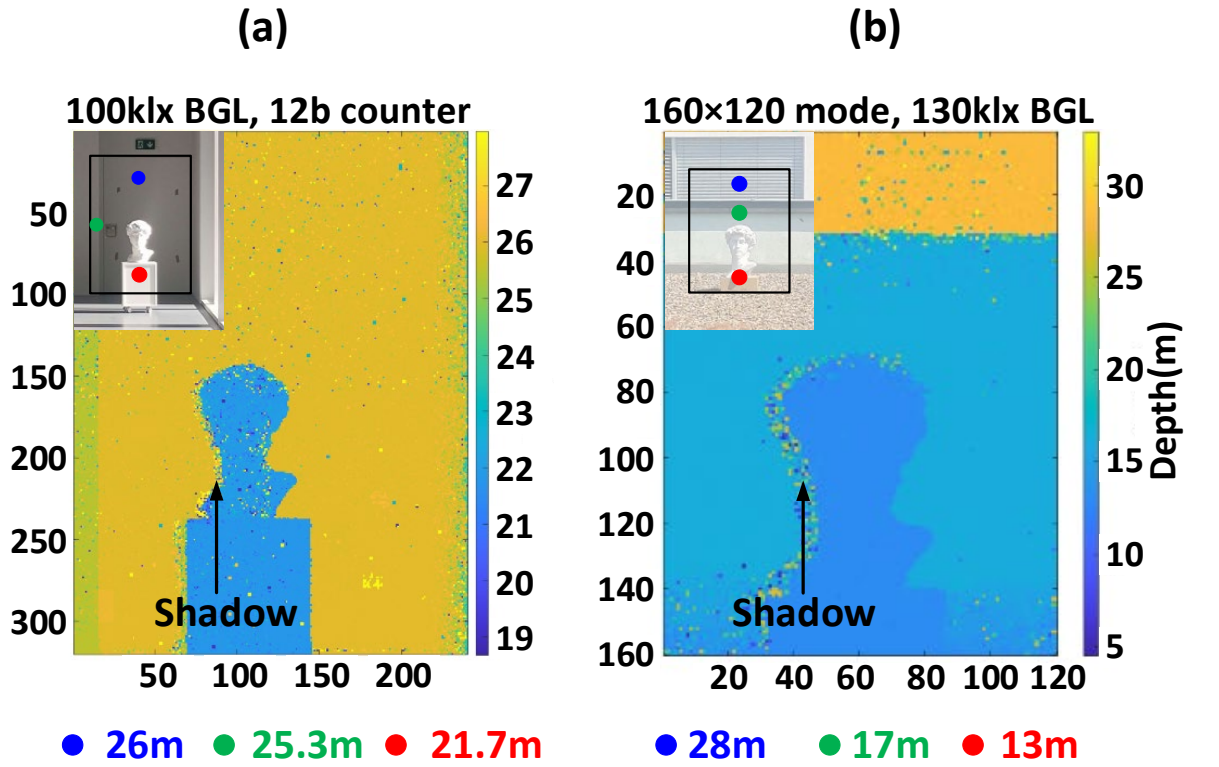

Fig. 11. Outdoor measurement results. (a) 3D image at 26m with 100klx BGL captured by the sensor. (b) 3D image at 28m with 130klx BGL captured by the sensor.

## IV. Conclusion

We have proposed a novel 320×240-pixel flash LiDAR sensor based on a SPAD pixel. To the best of our knowledge, this sensor achieves the largest imaging resolution and longest detection range among all reported flash LiDAR sensors. The proposed sensor demonstrates excellent performance in both short- and long-range measurements, indoors as well as outdoors. Fig. 12 compares the proposed flash LiDAR with other reported designs in terms of image resolution, detection range, and background-light suppression. Fig. 12(a) compares the detection range and Fig. 12(b) background-light immunity. The proposed sensor achieves a 320×240 resolution and a 108 m detection range. It achieves 130klx background light suppression when operating in the image resolution of 160×120 pixels. Under the same fabrication technology, the proposed LiDAR sensor also exhibits a significant advantage in pixel pitch. While [5] achieved a 7 μm pixel pitch, that sensor was fabricated in 3D-stacked technology and achieved only 10klx of background-light suppression capability. The design reported in [9] achieved 120klx of background-light suppression, but it featured only a 64×64-pixel array. The system in [11] offered a similar image resolution, but it was a scanning LiDAR rather than a flash architecture. Similarly, the work in [7] achieved the same image resolution, yet it is limited to indoor conditions with a maximum detection range of only 5.8m.

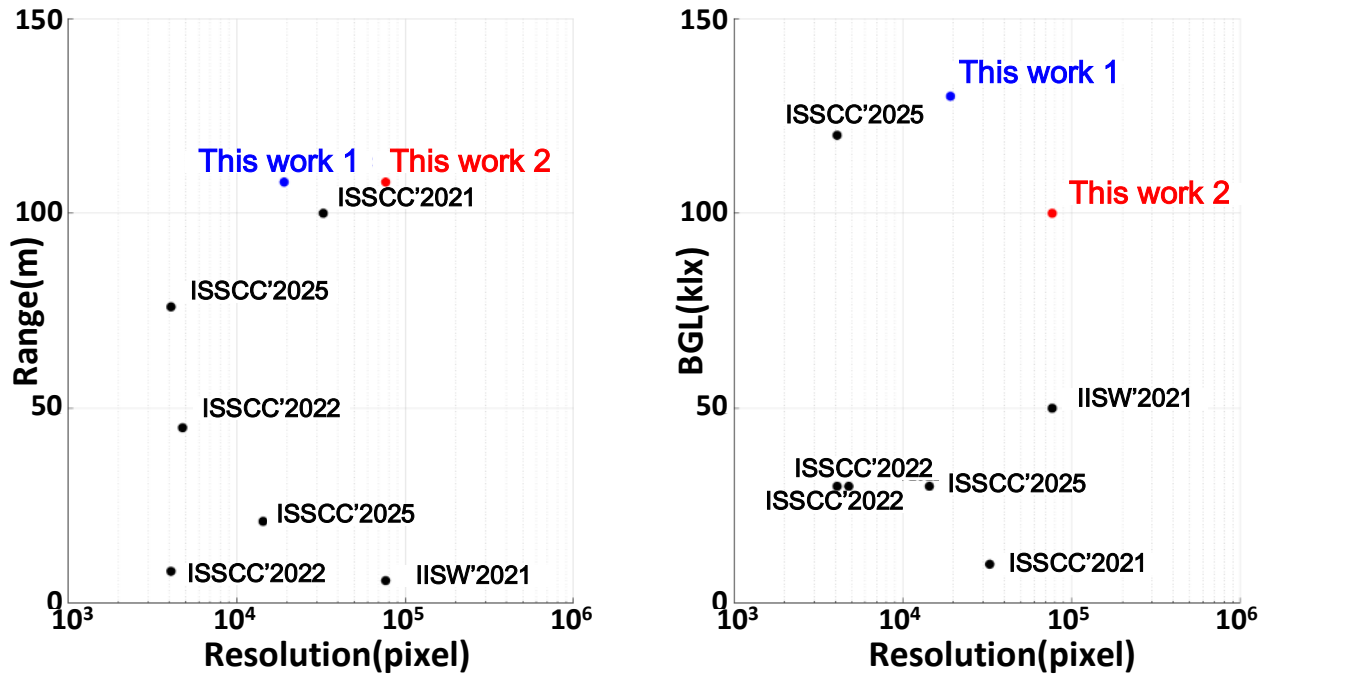

Fig. 12. Comparison summary [5, 7, 9-12]. (a). State-of-the-art comparison of detection range and pixel resolution in LiDAR sensors. (b). State-of-the-art comparison of BGL suppression and pixel resolution in LiDAR sensors.

Considering the detection range, spatial resolution, depth resolution, and background-light suppression capability, our sensor outperforms existing designs in terms of pixel resolution, measurement range, and background light suppression, showing its suitability for a wide range of industrial applications, such as autonomous driving, robotics, and 3D mapping.

Tab. 1 presents a comprehensive comparison, including two operating modes. Fig. 12 demonstrates that the imaging system outperforms existing designs in terms of pixel resolution, measurement range, and background light suppression.

TABLE I

PERFORMANCE SUMMARY AND COMPARISON WITH THE STATE-OF-THE-ART

| Parameters | This Work | | ISSCC'2025 Choi [9] | ISSCC'2025 Park [10] | ISSCC'2024 Kim [11] | ISSCC'2024 Han [12] | ISSCC'2021 Padmanabhan [5] | IISW'2021 Stoppa [7] |
|---|---|---|---|---|---|---|---|---|
| Process | **110nm** | | 110nm BSI | 90nm | 110nm | 110nm | 3D stacked 45nm BSI/ | 3D stacked 45nm/45nm |
| ToF type | **SPAD-based Direct ToF** | | SPAD-based Indirect ToF | SPAD-based Direct ToF | Scanning (Solid-State) | SPAD-based Direct ToF | SPAD-based Direct ToF | SPAD-based Direct ToF |
| Image Resolution | **320 × 240** | **160 × 120** | 64 × 64 | 160 × 90 | 320 × 240 | 160 × 120 | 256 × 128 | 320 × 240 |
| BGL suppression (klux) | **100** | **130** | 120 | 30 | 100 | / | 8 - 10 | 50 |
| Pixel Pitch (μm) | **23** | **2 × 23** | 32 | 52 | 16.5 × 14 | 35 | 7 | 12.5 |
| Max Distance (m) | **108** | **> 108** | 76 | 21 | 48 | 24 | 100 | 2 – 5.8 |
| Depth Precision (sigma) (cm/%) | **4.7cm@73m 27.4cm@108m** | **< 27.4cm @108m** | 6.4cm | 6cm | 8.8cm | 3.5cm | 7cm | < 0.5% |
| Illumination Power (mW) | **11.3** | | 5.96 | 60 | / | / | 5.0/1.5 | 93 |
| Illumination Wavelength (nm) | **780** | | 850 | 905 | 940 | 905 | 780 | 940 |
| Chip Size (mm²) | **8.2 × 6.3** | | 2.9 × 2.9 | 8.8 × 5.7 | 7.0 × 5.2 | 7.08 × 5.24 | 2.08 × 1.06 | / |
| Power Consumption(mW) | **298** | | 39.4 | 84 | / | 60 | 51.9 | < 1500 |
| Total Power Consumption /pixel (μW ) | **3.88** | **4 × 3.88** | 9.62 | 5.83 | / | 3.13 | 1.58 | / |